\documentstyle{article}

\def\be{\begin{equation}}
\def\ee{\end{equation}}
\def\bea{\begin{eqnarray}}
\def\eea{\end{eqnarray}}

\begin{document}

\section*{\center SEARCHES FOR STERILE NEUTRINOS AT REACTOR}

\Large
{\hspace{3.5 cm} L. Mikaelyan, V. Sinev}
\\
\vspace{0.5 cm}

\normalsize
{\it \center RRC "Kurchatov Institute", Kurchatov Sqr., 1, Moscow-123182, Russia\\
Talk given at RAS Nuclear Physics Department Conference, \\
\hspace{3 cm} Moscow, 16 - 20 November, 1998. }
\vspace{1em}\\
\righthyphenmin=3
We show that in a reactor disappearance oscillation experiment the sensitivity to the 
mixing parameter can be substantially improved for $(1 - 5\cdot10^{-4}) \ eV^{2}$ mass 
parameter range which includes both the LSND and the atmospheric neutrino oscillation 
regions. The objectives are: to search for the sterile neutrinos and to get a better 
understanding of the atmospheric neutrino oscillation mechanism. The parameters of the 
underground (600 m.w.e.) Krasnoyarsk reactor and data from the CHOOZ reactor oscillation 
experiment are used to estimate the effect and background rates.

\section{Introduction}
\large
It is now 40 years since neutrino oscillations and sterile neutrinos were first considered 
[1]. Recent study of atmospheric neutrinos [2] and analysis of solar neutrino observations 
[3] have provided evidence that neutrinos really oscillate and therefore are mixed and 
possess non zero masses. The relevant mass parameter ${\delta}m^{2}$ ranges have been 
localized to ${\delta}m^{2}_{atm}=(5\cdot10^{-4}-6\cdot10^{-3}) \ eV^{2},\\
{\delta}m^{2}_{sol}<5\cdot10^{-6} \ eV^{2}$
The available solar and atmospheric data do not restrict however the number of neutrino 
species involved. The sterile neutrinos can play a considerable and even a dominant part 
in the processes of mutual transformations responsible for the observed phenomena.

If the sterile neutrinos were found to exist in nature it would have a major impact on 
physics. We recall here the concept that all particles are duplicated to form a parallel 
mirror world suggested years ago to restore the fundamental symmetries lost as the result
of P and CP violation [4,5].

The ${\widetilde{\nu}}_{\mu}\rightarrow {\widetilde{\nu}}_{e}$ events at ${\delta}m^{2} > 0.1 \ eV^{2}$ 
reported by LSND collabora-tion [6] give evidence in favor of sterile neutrinos.
This has stimulated a great number of new studies on the theory and phenomenology of sterile 
neutrinos (see publications [7,8,9] and references therein).

Here we consider a reactor disappearance oscillation experiment and show that the sensitivity 
to the mixing parameter can be greatly improved for $(1 - 5\cdot10^{-4}) \ eV^{2}$ mass 
parameter range which includes both the LSND events [6] and atmospheric neutrino 
oscilla-tion regions. The main objectives we pursue are: \\
- to search for the sterile neutrinos and \\
- to get a better understanding of the atmospheric neutrino oscillations.

We plan to use relatively small neutrino detectors: the neutrino target (liquid scintillator) 
will not exceed by mass a few percent of the targets used in the most part of the running 
or projected scintillation neutrino detectors.

The parameters of the underground (600 m.w.e.) Krasnoyarsk reactor and data from the CHOOZ 
reactor oscillation experiment are used to estimate the effect and background rates.

\section{SIGNATURE FOR STERILE NEUTRINOS}

The electron neutrino flavor state ${\nu}_{e}$ is connected with the mass eigenstates 
${\nu}_{i}$ by the relation
\begin{equation}
{\nu}_{e}=\sum_{i}{U_{ei}{\nu}_{i}}
\end{equation}
The survival probability $P({\nu}_{e}\rightarrow {\nu}_{e})$ of ${\nu}_{e}$ at the distance 
R (in m) from the source is given by the expression 
\begin{equation}
P({\nu}_{e}\rightarrow {\nu}_{e})=1 - 4\sum{U^{2}_{ej}U^{2}_{ei}sin^{2}(1.27{\delta}m^{2}_{ji}R/E)}, \ (j > i),
\end{equation}
where E is the neutrino energy in MeV and ${\delta}m^{2}_{ji}=m^{2}_{j}-m^{2}_{i}$ are the mass
parameters in $eV^{2}$  ($m_{j} > m_{i}$ for $j > i$).

In the three active neutrino mixing scheme ${\nu}_{\alpha}, \ ({\alpha}=e, {\mu}, {\tau})$ there
are three mass states ${\nu}_{i}$, (i = 1,2,3) and three mass parameters
${\delta}m^{2}_{21}=m^{2}_{2}-m^{2}_{1}$, \ ${\delta}m^{2}_{31}=m^{2}_{3}-m^{2}_{1}$ \quad and 
${\delta}m^{2}_{32}=m^{2}_{3}-m^{2}_{2}$, of which two are independent 
(${\delta}m^{2}_{32}={\delta}m^{2}_{31}-{\delta}m^{2}_{21}$).

All potential vacancies available in 3 neutrino mixing scheme are however already 
occupied by the solar and atmospheric candidates:
(${\delta}m^{2}_{sol} < 5\cdot10^{-6} \ eV^{2}$, ${\delta}m^{2}_{atm} = 5\cdot10^{-4}-6\cdot10^{-3} \ eV^{2}$
and ${\delta}m^{2}_{atm}-{\delta}m^{2}_{sol} \approx {\delta}m^{2}_{atm}$.

We can therefore conclude that if in the reactor oscillation experi-ment a disappearance 
effect is found at ${\delta}m^{2} < {\delta}m^{2}_{atm}$ it would mean that at least four 
mass states and one sterile neutrino must exist in nature. In the three active and three 
sterile neutrino oscillation scenario there are six mass states and 15 mass parameters.
This increases chances to find some of the relevant transitions in
the ${\delta}m^{2}$  range we suggest to inspect.

\section{EXPERIMENT}

3.1  We consider two identical liquid  scintillation spectrometers, stati-oned at the 
distances $R_{1}$ and $R_{2}$ from the reactor antineutrino source. The positron energy 
spectra $S(E_{e})$ are measured simultaneous-ly via the inverse beta decay reaction 
$$
{\widetilde {\nu}}_{e} + p \to e^{+} + n \quad \mbox{with} \quad 
E_{e} = E - 1.80 \quad \mbox{MeV.}
$$
The usual delayed coincidence technic between the prompt $e^{+}$ signal (boosted by the 
annihilation gamma rays) and the signal from the neutron capture 2.2 MeV gamma 
(no Gd is added to the scintillator).

Small deviations of the ratio $S_{1}/S_{2}$
\begin{equation}
S_{1}/S_{2} = C\cdot(1 - sin^{2}2{\theta}sin^{2}{\phi}_{1})\cdot(1 - sin^{2}2{\theta}sin^{2}{\phi}_{2})^{-1} ,
\end{equation}
from a constant value are searched for the oscillation effects (${\phi}_{1,2}$ stands for 
$1.27{\delta}m^{2}R_{1,2}/E$).

No knowledge of the constant C in Eq. (3) is needed for this analysis so that the details 
of the geometry, ratio of the target masses etc are excluded from the consideration.

3.2  A simplified version of the BOREXINO detector composition [10] is chosen for the 
design of the spectrometers (Fig.1). The target in the center of the detector (mineral 
oil + PPO) is viewed by  the PMT's ($\sim 15\%$ coverage, 150 photoelectrons/MeV) through 
sufficiently thick layer of the oil.

The expected positron energy spectrum is shown in the Fid.2 by the solid line.

The search for oscillations in the mass parameter range as wide as $(1 - 5\cdot10^{-4}) \ eV^{2}$
can be performed in two steps. First, the measurements are done at the distaces of 20 m 
- 100m with small detectors, then larger detectors are used to cover the 100 m - 1 km 
region. The projected target masses and expected neutrino detection rates can be
seen in the Table 1.
\normalsize
\begin{table}[h]
\caption{Expected reactor ${\widetilde {\nu}}_{e}$ detection rates $N(e^{+},n)$/day in this
experiment and in the BOREXINO [11] and KamLAND [12] projects\label{tab:exp}}
\vspace{0.4cm}
\begin{center}
\begin{tabular}{|c|c|c|c|c|}
\hline
Detector &\multicolumn{2}{|c|}{THIS EXPERIMENT}&BOREXINO & KamLAND\\
\cline{2-3}
 Target mass & 3.4 t & 50t & 300 t & 1000 t \\
\hline
Detector-  &     &     &     &    \\
reactor dist. &  20 m  100 m & 100 m 300 m 1 km & 800 km & 160 km  \\
\hline
$N(e^{+},n)$/day &10 000 \ 400  & 7 000 \  750 \ 70 &   0.08  &   2 \\
\hline
\end{tabular}
\end{center}
\end{table}

\large
The rates are found with the computed ${\widetilde {\nu}}_{e}$ detection efficiencies
${\epsilon} = 70\%$ for the smaller detector and ${\epsilon} = 85\%$ for the bigger one.

3.3  We consider two components of the background. \\
- The natural radioactivity of the external and internal materials such as the walls of the room, the PMT's, the
            mineral oil etc. \\
-  All sort of effects produced by cosmic muons.

The neutrino detection rate expected for the 1 km position (see Tab.1) when scaled to the 
same target mass is ${\sim}$ three orders of magnitude higher than in the KamLAND and the 
BOREXINO projects. So we hope that no extraordinary efforts are needed to keep the 
background due to the radioactivity at a sufficiently low level.

To estimate the muon induced component we use the absolute value of the background 
measured in the CHOOZ experiment [13]:
\quad ${\sim}$ 1 per day per 5 ton target at the depth of 300 mwe /CHOOZ/.

At Krasnoyarsk with the depth of 600 mwe the ${\sim}$ 5 times lower level of the background is 
expected, which indicates that in the \\ 1 km position the effect to background ratio not 
worse than 20 : 1 can be hoped for.

3.4  Calibrations of the detector are of crucial importance. The difference between the 
energy scales of the two detectors which is difficult to avoid can give additional 
modulation to the ratio of the spectra (3) and thus imitate the oscillation effect.

The difference can be measured and relevant corrections found by systematic intercomparison 
of the scales in many energy points using the sources of ${\gamma}$-rays listed in the Table 2.
\begin{table}[h]
\caption{Sources of ${\gamma}$-rays for calibrations\label{tab:gam}}
\vspace{0.4cm}
\begin{center}
\begin{tabular}{|c|c|c|c|c|c|c|}
\hline
 Source & $^{137}Cs$ & $^{65}Zn$ & $^{22}Na$ & (n,${\gamma}$) & $^{60}Co$ & $^{24}Na$ \\
\hline
 Energy, & 0.662  &  1.115 &  1.022 &  2.23   & 2.505  &  4.122 \\
  MeV    &        &        &  2.296 &         &        &        \\
\hline
\end{tabular}
\end{center}
\end{table}

In addition an overall tests can be done with the use of the $^{252}Cf$ spontaneous 
fission source which can produce in the detector a broad spectrum due to prompt ${\gamma}$-
rays and neutron recoils (see the broken line in the Fig.2).

\section{ EXPECTED RESULTS}

4.1 Expected $90\%$ CL exclusion contour is shown in Fig.3. It can be seen that in the 
(0.7 - 0.007) \ $eV^{2}$ mass range the sensitivity to the mixing angle better than 
0.01 can be reached. For ${\delta}m^{2} > 1 \ eV^{2}$ the sensitivity is decreasing 
because of the finite size of the ${\widetilde {\nu}}_{e}$ source while the small 
${\delta}m^{2}$ region is influenced by poor statistics at the 1 km position.

4.2 It is highly desirable to get more information on the ${\delta}m^{2}_{atm}$ region. 
According to the 1998 y. SUPERKAMIOKANDE publication [2] the allowed mass parameters for 
atmospheric oscillations have shifted towards smaller ${\delta}m^{2}$ values relative to the 
KAMIOKANDE data. The results from CHOOZ leave now a considerable freedom for the 
${\nu}_{e}\rightarrow {\nu}_{e}$ atmospheric neutrino oscillations (see Fig.3)...

Eqs.(1,2) show that the ${\widetilde {\nu}}_{e}$ disappearance probabilities depend only 
on its mass content. So some information on the issue can be expected from the data.

In three neutrino oscillation model at $\sim$ 1 km from the reactor ${\widetilde {\nu}}_{e}$ 
can oscillate only if the mass state ${\nu}_{3}$ contributes to the electron neutrino 
flavor state: using Eq.(2) we obtain by direct calculation:
\begin{equation}
sin^{2} 2{\theta} = 4U^{2}_{e3}(1 - U^{2}_{e3}) \approx 4U^{2}_{e3}
\end{equation}
For ${\delta}m^{2}$ = 0.002 $eV^{2}$ (taken as the most probable value) the ${\nu}_{3}$
contribution as small as $U^{2}_{e3} > 5\cdot10^{-3}$ can be observed in the experiment 
under consideration.

In  six neutrino oscillation model $sin^{2} 2{\theta}$ will depend on mass spectrum in a 
more specific manner. Here are two examples. Suppose that holds the mass hierarchy
\begin{equation}
m^{2}_{6} \gg m^{2}_{5} \gg m^{2}_{4} ...\gg m^{2}_{1} \ .
\end{equation}
In this case the 15 mass parameters are clustered in 5 tight groups:
$$
{\delta}m^{2}_{i} \approx m^{2}_{i} \qquad (i = 2,3...6) \eqno (5')
$$
If we now assume for a moment that there exist no mass parameter greater then ${\delta}m^{2}_{atm}$ 
( $\approx m^{2}_{6}$ ) then instead of Eq.(4) we find:
\begin{equation}
sin^{2} 2{\theta} = 4U^{2}_{e6}(1 - U^{2}_{e6})
\end{equation}
If however we consider ${\delta}m^{2}_{LSND} \approx m^{2}_{6}$ as the maximal mass parameter and use 
${\delta}m^{2}_{atm} \approx m^{2}_{5}$  then the reactor neutrino survival probability can be expressed as:
\bea
P({\nu}_{e}\rightarrow {\nu}_{e}) = 1 - 4U^{2}_{e6}(1 - U^{2}_{e6})sin^{2}(1.27{\delta}m^{2}_{LSND}R/E)\nonumber\\
\quad - 4U^{2}_{e5}(1 - U^{2}_{e5}- U^{2}_{e6})sin^{2}(1.27{\delta}m^{2}_{atm}R/E)
\eea

Some other representations can be found in Refs.[9, 15].

Note also that in the case of six neutrinos some of the mass parameters get in the solar 
neutrino region ${\delta}m^{2}_{sol} < 6\cdot10^{-6} \ eV^{2}$  which can probably 
influence some results of the analysis given in Ref.[3].

\section{CONCLUSIONS}

We have shown that with reactor neutrinos the sensitivity to the neutrino mixing can 
considerably be improved for $(1 - 5\cdot10^{-4}) \ eV^{2}$ mass parameter range. This 
gives opportunities for systematic searches for the sterile neutrinos and opens promising 
perspectives for a better understanding of the atmospheric neutrino oscillations.

\section*{Acknowledgments}

We thank our colleagues A.Etenko, Yu.Kozlov, V.Kopeikin, I.Machulin, V.Martemyanov, 
M.Skorokhvatov, S.Sukhotin and V.Vyrodov for useful criticism. Fruitful discussions with 
E.Akhmedov, S.Bilenky, L.Okun and A. Smirnov are greatly appreciated by L.M.

This study is supported by RFBR, grant N 96-15-96640.


\end{document}